\documentstyle[12pt]{article}
\begin{document}

\begin{flushright}
UI-NTH-9306
\end{flushright}

\begin{center}
{\LARGE\bf Comment on
\\
\vspace*{.1cm}
``Comparison of potential models with the $pp$
scattering data below 350 MeV"}
\\
\vspace*{.5cm}
{\large\sc G. Q. Li and R. Machleidt}
\\
{\it Department of Physics, University of Idaho,
 Moscow, Idaho 83843, U.S.A.}
\\
\today
\end{center}

\vspace*{.5cm}

\begin{abstract}
We point out two flaws in the recent test of nucleon-nucleon (NN)
potentials conducted by Stoks and de Swart. First, in some cases,
the neutron-proton ($np$) version of an NN
potential was compared to the proton-proton ($pp$) data, which
is improper and yields (large) $\chi^2$ that are essentially meaningless.
Second, for a proper test of the quantitative nature of a NN potential,
it is insufficient to compare to $pp$ data only, since this leaves the T=0
potential untested.
Thus, it can happen that the $pp$ version of a potential predicts the
$pp$ data accurately, while the $np$ version of that same potential
is poor in $np$
(where also the T=0 potential is involved).
An example for this is the Nijmegen potential,
which predicts the $pp$ data well with a $\chi^2$/datum of 2.0,
but yields a $\chi^2$/datum of 6.5 in $np$.
\end{abstract}

%\pagebreak
\vspace*{.5cm}

In a recent paper \cite{SS93}, Stoks and de Swart
compare some nucleon-nucleon (NN) potential models with the
proton-proton ({\it pp}) scattering data below 350 MeV.
The general purpose of their study is  to test the quantitative nature
of these NN models, since this is important when these potentials are applied
in ``three-nucleon elastic scattering, few-nucleon bound-states, and nuclear
matter calculations''~\cite{SS93}.
Moreover, the authors stress ``that one has to be very careful
in drawing conclusions
regarding the importance or unimportance of, e.~g., three-nucleon
forces in many-body
calculations, when these conclusions are only based on calculations
where the NN interaction is represented by an NN potential model which
cannot even adequately describe the two-nucleon scattering data''~\cite{SS93}.

We strongly agree with the authors of Ref.~\cite{SS93} that a
comprehensive and reliable knowledge
of the quantitative nature of an NN potential is important to
properly assess results based upon the potential.
However, we are concerned
that the information provided by the Nijmegen group
in their study
is, in part, incomplete and may, in part, be misleading to some non-experts.
Essentially, we see two flaws in the Nijmegen investigation.
First, in some cases, the Nijmegen group considers neutron-proton ($np$)
potentials to calculate the $\chi^2$ for the fit of the proton-proton
($pp$) data; this is improper and yields huge $\chi^2$ values that are
basically meaningless.
Second, the Nijmegen analysis is restricted to a comparison
with the $pp$ data only.
This leaves the T=0 part of the NN potential untested. However, this
part of the potential is very important in, e.~g., calculations of few-nucleon
bound-states and three-nucleon scattering, which are
part of the motivation for the Nijmegen study (see quotes above).

We will use the rest of this Comment to explain our two points of
concern in more
detail.

It is very important that a $\chi^2$ is calculated properly.
The most important rule here is:
{\it A $pp$ potential must only be confronted with
$pp$ data, while a $np$ potential must only be confronted with $np$
data.}
Though this rule is obvious, it has
been violated
in Ref. \cite{SS93}
in the case of the Argonne
\cite{WSA84}  and the ``Bonn87'' potentials \cite{MHE87}, which are $np$
potentials by construction.
Let us briefly explain why this rule is so important.
At low energies, NN scattering takes place mainly in $S$ wave.
There is well-known charge-dependence in the $^1S_0$ state and
the electromagnetic effects are very large
in low energy $pp$ scattering.
Thus, $np$ and $pp$ differ here substantially.
Moreover, there exist very accurate $pp$ cross section data
at low energies.
Consequently, if (improperly)
a $np$ potential is applied to $pp$ scattering,
a very large $\chi^2$ is obtained. However, this large $\chi^2$
has nothing to do with the quality of the $np$ potential;
it simply reflects the fact
that charge-dependence is important
and that the $pp$ data carry a very small error at low energies.

To give an example:
When  the $np$
versions of the Argonne~\cite{WSA84} and
``Bonn87''~\cite{MHE87,foot1} potentials
are (improperly)
confronted with the $pp$ data, a $\chi^2$/datum of 824 and 641,
respectively,
is obtained for the energy range 0--350 MeV~\cite{Swa89};
for 2--350 MeV the $\chi^2$/datum are 7.1 and 13, respectively
(cf.\ Table II of Ref. \cite{SS93}).
However, if (properly) the $pp$ version of the Bonn potential
is confronted with the $pp$ data, a $\chi^2$/datum of 1.9 is obtained
(cf.\ ``Bonn89'' in Table II of Ref.~\cite{SS93})~\cite{foot2}.
It is now important to notice that
the change in the potential, that brings about
this large change in the $\chi^2$, is minimal.
The main effect comes from the $^1S_0$. A $np$ potential is fitted
to the $np$ value for the singlet scattering length.
Now, if one wants to construct
a $pp$ potential from this, one has to do essentially only two things:
The Coulomb force has to be included and
the singlet scattering length
has to be readjusted to its $pp$ value. Since the scattering length
of an almost bound state is a super-sensitive quantity,
this is achieved by a very small
change of one of the fit parameters; for example, a change of the
$\sigma$ coupling constant by as little as 1\%.
This is all that needs to be done;
this
changes the
$\chi^2$/datum from 641 to 2.
It shows in a clear way how misleading $\chi^2$ can be if
the reader is
not familiar with the field.

The physically more interesting and relevant question is
what difference it makes in microscopic nuclear structure calculations
whether the NN potential used is adjusted to $pp$ or $np$.
To give two examples: In nuclear matter, the binding energy per nucleon
at normal nuclear matter density
comes out 0.61 MeV smaller for a $pp$ potential
as compared to an $np$ potential~\cite{LM93}.
The correct charge-dependent calculation is 0.27 MeV above the $pp$ value
for the binding energy.
This must be compared to the total nuclear matter
binding energy per nucleon of 16 MeV.
With regard to the large uncertainties that nuclear matter theory is
beset with, charge-dependence is a negligible effect in nuclear matter
at the present time.

The situation is different for the three-nucleon problem, where rigorous
Faddeev
 calculations are performed. Here, $np$ potentials predict about 0.3 MeV more
binding energy for the triton than $pp$ potentials. The correct
charge-dependent
calculation is 0.1 MeV above the $pp$ result. Since the gap between
predictions from two-body forces and the experimental value for the
triton binding of 8.48 MeV is between 0.2 and 1 MeV, charge-dependence
is important in three-nucleon bound state calculations.
For three-nucleon scattering (e.~g., $n-d$ elastic and
breakup) charge-dependence may even be crucial,
as shown by the Bochum group~\cite{WGK91,WG91}.

In summary, one should in general carefully distinguish between
the $np$ and the $pp$ version of an NN potential. This distinction
is absolutely crucial for the calculation of the $\chi^2$ of the fit
of the NN scattering data. In most nuclear structure calculations,
charge-dependence is not important at the present time; a remarkable
exception occurs, however, in
`exact' few-nucleon calculations, for which charge-dependence is crucial
in some cases.

The second point which we would like to explain in this Comment
is the fact that
{\it for testing
the quantitative nature of a  NN potential, it is insufficient
to make a comparison  with $pp$ data only.}
Proton-proton states are T=1 (where T denotes the total isospin of the
two-nucleon system) and, thus, a comparison with the $pp$
data tests only the T=1 potential. However, there is also the
T=0 potential, which is equally
important for applications in few-nucleon physics and nuclear
structure; in fact, one may well argue that the T=0 potential is
more crucial, since the important $^3S_1$ state is T=0.
This isoscalar potential is tested only in a comparison with $np$ data
(in which both isospin states are involved).

Since T=0 is excluded from $pp$, it may happen that the $pp$ version
of a potential is very
successful in $pp$,
while the $np$ version of that same potential fails in $np$.
To illustrate this point, we show in Table I the $\chi^2$/datum for three
modern potentials which describe the $pp$ data about equally well
($\chi^2$/datum $\approx 2$ in all three cases, in agreement with the
findings of Ref.~\cite{SS93})~\cite{foot3}.
In the second row of Table~I, the $\chi^2$/datum for the fit of the $np$ data
(using the $np$ version of the potentials)
is given. It is clearly seen that, in some cases, this $\chi^2$ is
substantially different (factor 2-3 larger) from $pp$.
In the case of the Paris potential (and in part for the Nijmegen potential)
the large $np$ $\chi^2$ is
essentially due to the fact that the $np$ total cross sections
($\sigma_{tot}$) are predicted too large, which in turn is due to
too large $^3D_2$ phase shifts (see Fig.~1b).

While the T=1 phase-shift predictions by modern potentials
(e.~g., Nijmegen, Paris, and Bonn) are so close that in conventional graphs
they are almost indistinguishable, the situation is very different for
T=0. To demonstrate this point, we show in Fig.~1 some T=0
phase shifts.
Clearly there are substantial differences among the predictions by the
three models considered.
There are corresponding differences in the predictions for $np$ spin
observables of which we show two very recent measurements in Fig.~2 and 3.
Differences in the predictions for the spin-correlation parameter
$A_{yy}$ (Fig.~2)  can be clearly traced to
$^3D_2$ and $^3D_3$, while in $A_{zz}$ (Fig.~3) the differences in
the predictions for the $^1P_1$ phase shifts show up.

In summary, the largest differences between modern NN potentials occur
in the T=0 states. Thus, a pure $pp$ investigation (restricted to T=1)
misses important information that may have serious implications for
calculations of few-nucleon bound-states, three-nucleon scattering,
and other applications.

\vskip .5cm
One of the authors (R. M.) would like to thank R. B. Wiringa
for interesting discussions and H. Willmes for suggestions on the manuscript.
This work was supported in part by the U.S. National Science Foundation
under Grant No. PHY-9211607,
and by the Idaho State Board of Education.

%\newpage

\pagebreak

\noindent
{\bf Table~1.} $\chi^2$/datum for fit of world NN data by some
current models for the NN interaction.
\begin{center}
\begin{tabular}{lccc}
\hline\hline
            & Nijmegen~\cite{NRS78} & Paris~\cite{Lac80} &
Bonn full model$^a$\\
\hline
all $pp$ data & 2.06 & 2.31 & 1.94 \\
all $np$ data & 6.53 & 4.35 & 1.88 \\
($np$ without $\sigma_{tot}$)&(3.83)& (1.98) & (1.89)  \\
all $pp$ and $np$&5.12              &  3.71  &  1.90   \\
\hline\hline
\end{tabular}
\end{center}
The $\chi^2$/datum  are obtained
from the computer software SAID of R. A. Arndt and L. D. Roper
(VPI\&SU)~\cite{SAID}.
The world NN data set in the range 10--300 MeV as of September 1992
is used~\cite{foot3}; it includes 1070 data for $pp$, 2158 data
for $np$ without total cross sections
($\sigma_{tot}$), and 2322 data for $np$ with $\sigma_{tot}$.
\\
$^a$ The $np$ version of the Bonn full model is published
in the original paper~\cite{MHE87}, the $pp$ version can be found in
Refs.~\cite{HH89,ML93}; the phase shifts for both $np$ and $pp$ are
available from SAID~\cite{SAID}.

\pagebreak

\begin{center}
\large FIGURE CAPTIONS
\end{center}

\noindent
{\bf Figure 1.}
Some
T=0 phase shifts. {\bf (a)} $^1P_1$,
{\bf (b)} $^3D_2$, and {\bf (c)} $^3D_3$. Predictions are shown by
the Nijmegen potential~\cite{NRS78} (dotted line),
Paris potential~\cite{Lac80} (dashed),
and the Bonn full model~\cite{MHE87} (solid line).
The solid dots represent the energy-independent phase shift analysis by
Arndt {\it et al.}~\cite{AHR87}.
\\ \\
{\bf Figure 2.} Neutron-proton spin correlation parameter $A_{yy}$
at 181 MeV.
Predictions by the Nijmegen potential~\cite{NRS78} (dotted line),
Paris potential~\cite{Lac80} (dashed), and
Bonn full model~\cite{MHE87} (solid line)
are compared with the data (solid dots)
from Indiana~\cite{Indiana}.
The $\chi^2/$datum for the fit of these data is
54.4 for Nijmegen, 3.22 for Paris, and 1.78 for Bonn~\cite{SAID}.
The experimental error bars include only systematics and statistics;
there is also a scale error of $\pm 8$\%.
In the calculations of the $\chi^2$, all three errors have been
taken into account~\cite{SAID}.
\\ \\
{\bf Figure 3.} Neutron-proton spin correlation parameter $A_{zz}$
at 67.5 MeV.
Predictions by the Nijmegen potential~\cite{NRS78} (dotted line),
Paris potential~\cite{Lac80} (dashed), and
Bonn full model~\cite{MHE87} (solid line)
are compared with the data (solid dots)
taken by the Basel group~\cite{Basel}.
The $\chi^2/$datum for the fit of these data is
47.7 for Nijmegen, 1.6 for Paris, and 1.2 for Bonn~\cite{SAID}.
The experimental error bars include only systematics and statistics;
there is also a normalization uncertainty of $\pm 6$\%.
In the $\chi^2$ calculations, all three errors have been
taken into account~\cite{SAID}.

\newpage

\vspace*{3cm}

The figures, which are crucial for a proper understanding of this Comment,
are available upon request from
\begin{center}
{\sc machleid@tamaluit.phys.uidaho.edu}
\end{center}
Please, include your FAX-number with your request.


\begin{thebibliography}{99}
\bibitem{foot3}
The $\chi^2$ given in our Table~I are calculated for the range 10--300 MeV
(in terms of the kinetic energy of the incident nucleon in the laboratory
system). We find this energy range more appropriate than the range
0--350 MeV, for the following reasons.
For energies below 10 MeV, the effective range expansion applies and
the quantitative nature of a potential can be tested by checking how well
it reproduces the empirical effective range parameters.
This is physically more reasonable than calculating (possibly, huge)
$\chi^2$.
The low energy parameters as predicted by the Nijmegen, Paris, and Bonn
potential are compared to the empirical values
in Table~2 of Ref.~\cite{ML93} and we will not
repeat this here. Concerning the upper end of the energy range to be
considered for testing real potentials, one has to keep in mind that
pion production starts around 280--290 MeV. This implies that the real
potential concept is strictly speaking wrong above these energies.
It may be o.~k. to stretch the limit by a few MeV up to 300 MeV, but not
beyond that. Not surprisingly, for most potentials,
the $np$ $\chi^2$/datum is already very
bad for the range 300--350 MeV, e.~g.,  18.0, 7.2, and 6.2 for
the Nijmegen, Paris, and Bonn potentials, respectively.
Thus, including this energy range will increase the over-all $\chi^2$
substantially, due to contributions that are rather irrelevant.
\end{thebibliography}
\end{document}